\documentstyle[11pt,amsmath,amssymb,graphicx]{article}

\def\be{\begin{eqnarray}}
\def\ee{\end{eqnarray}}

\def\p{\partial}
\def\tr{{\rm tr}\,}

\hoffset= - 1.0in         

\title{{\bf Brezin-Gross-Witten model as "pure gauge" limit of Selberg integrals}
\vspace{.2cm}}
\author{{\bf A.Mironov}\footnote{ {\small {\it
Lebedev Physics Institute} and {\it ITEP, Moscow, Russia}};
mironov@itep.ru; mironov@lpi.ru}, {\bf A.Morozov}\thanks{{\small
{\it ITEP, Moscow, Russia} and {\it Laboratoire de Mathematiques et
Physique Theorique, CNRS-UMR 6083, Universite Francois Rabelais de
Tours, France}}; morozov@itep.ru} \ and {\bf
Sh.Shakirov}\thanks{{\small {\it ITEP, Moscow, Russia} and {\it
MIPT, Dolgoprudny, Russia}}; shakirov@itep.ru}\date{ }}

\begin{document}

\maketitle

\vspace{-6.0cm}

\begin{center}
\hfill FIAN/TD-08/10\\
\hfill ITEP/TH-39/10\\
\end{center}

\vspace{4cm}

\begin{abstract}
The AGT relation identifies the Nekrasov functions for various
${\cal N} = 2$ SUSY gauge theories with the 2d conformal blocks,
which possess explicit Dotsenko-Fateev matrix model
($\beta$-ensemble) representations the latter being polylinear
combinations of Selberg integrals. The "pure gauge" limit of these
matrix models is, however, a non-trivial multiscaling large-$N$
limit, which requires a separate investigation. We show that in this
pure gauge limit the Selberg integrals turn into averages in a
Brezin-Gross-Witten (BGW) model. Thus, the Nekrasov function for
pure $SU(2)$ theory acquires a form very much reminiscent of the AMM
decomposition formula for some model $X$ into a pair of the BGW
models. At the same time, $X$, which still has to be found, is the
pure gauge limit of the elliptic Selberg integral. Presumably, it is
again a BGW model, only in the Dijkgraaf-Vafa double cut phase.
\end{abstract}


\section{Introduction}

\paragraph{The pure gauge limit.} The AGT relation \cite{AGTfirst}-\cite{AGTlast}
is an explicit formulation of duality between the $2d$ and $4d$
descriptions of conformal $6d$ theory of self-dual 2-forms,
compactified on a Riemann surface \cite{Gai}. The theory of the
corresponding $M5$-brane is long known to be related to
integrability theory \cite{M5}, but explicit route from
integrability to the AGT relation still remains a mystery.

A promising approach to origins of the AGT relations is through
their reformulation as relations between matrix models \cite{UFN3}
and Seiberg-Witten theory \cite{SW,SWint}, see \cite{MMSproof} for a
concise review of this idea (which is a new application of the
topological recursion \cite{AMMf}-\cite{AMMbgw}).

Surprisingly or not, despite obvious conceptual advantages of such
an approach, some simple properties of original AGT relations are
not so easy to describe in the matrix-model reformulation. A typical
example is the "pure gauge" limit (PGL), where the dimensional
transmutation takes place and the conformal invariance gets broken.
In the matrix model formulation, this corresponds to a non-trivial
double-scaling large $N$ limit of the relevant matrix models, which
will be the subject of the present paper.

In this paper, we concentrate on the simple case of pure $SU(2)$
Nekrasov function, which (in the context of the AGT relations)
arises as the PGL of either the $4$-point conformal block
$B^{(0)}(\Delta_1,\Delta_2,\Delta_3,\Delta_4;\Delta,c|x)$ on
sphere \cite{PGL0} or the $1$-point conformal block
$B^{(1)}(\Delta_{ext};\Delta,c|q)$ on torus (where $q=e^{2\pi
i\tau}$ and $\tau$ is the modulus) \cite{PGL1,MMMsurop}:

\be
B_*(\Delta|\Lambda) \ = \
\lim_{
\stackrel{
\stackrel{\Delta_1,\Delta_2,\Delta_3,\Delta_4\rightarrow \infty, \ x\rightarrow 0}{}
}
{x(\Delta_2-\Delta_1)(\Delta_3-\Delta_4) \equiv \Lambda^4}
}
\Big[ B^{(0)}(\Delta_1,\Delta_2,\Delta_3,\Delta_4;\Delta,c|x) \Big] =
\lim_{
\stackrel{
\stackrel{\Delta_{ext}\rightarrow \infty, \ q\rightarrow 0}{}
}
{q\Delta_{ext}^2 \equiv \Lambda^4}
}
\Big[ B^{(1)}(\Delta_{ext};\Delta,c|q) \Big]
\label{pglB}
\ee
\smallskip\\
Our aim is to study this pure gauge limit at the level of matrix models.

\paragraph{PGL in the matrix model formulation.} Fortunately, matrix model
(i.e. the Dotsenko-Fateev-like $\beta$-ensemble) representations are
already known both for $B^{(0)}$ \cite{B0mamo,MMMsurop} and for $B^{(1)}$
\cite{B1mamo}. The first one is represented \cite{Ito} as an AMM
decomposition \cite{AMMmer,AMMbgw} into two spherical Selberg
integrals
\be
B^{(0)}(x) = \exp\left( -\dfrac{2}{\beta} \sum\limits_{k =
1}^{\infty} \dfrac{x^k}{k} \left[ \dfrac{v_+}{2} +
\dfrac{\partial}{\partial t^{+}_k} \right] \left[
\dfrac{v_-}{2} + \dfrac{\partial}{\partial t^{-}_k} \right]
\right) Z_{S}^{(0)} \big( u_+, v_+, N_+ \big| t^{+}  \big)
Z_{S}^{(0)} \big(u_-, v_-, N_- \big| t^{-}  \big) \Big|_{t=0}
\label{decoB0}
\ee
while the second one is the single, but elliptic
(toric) Selberg integral
\be
B^{(1)}(q) = \prod_{p = 1}^{\infty} (1
- q^p)^{\nu} \ Z^{(1)}_S\big( A, N \big| q \big) \label{decoB1}
\ee
Then, eqs.(\ref{pglB}), (\ref{decoB0}) and (\ref{decoB1}) imply that
\be
B_*(\Delta|\Lambda) = \exp\left( -\dfrac{2}{\beta} \sum\limits_{k =
1}^{\infty} \dfrac{\Lambda^{4k}}{k} \dfrac{\partial^2}{\partial
t^{+}_k \partial t^{-}_k} \right) Z_{*+}^{(0)}\big( t^{+}\big)
Z_{*-}^{(0)}\big( t^{-} \big)\Big|_{t=0} = Z_*^{(1)}(A|\Lambda)
\label{decoB*}
\ee
and this is the straightforward matrix model
formulation of eq.(\ref{pglB}). The partition functions $Z_{*\pm}^{(0)}$
and $Z_*^{(1)}$ denote here what we call the PGL of the
corresponding Selberg integrals. To describe this PGL, explicit
expressions for the Selberg integrals are needed. Note that the second
equality in (\ref{decoB*}) automatically provides an AMM
decomposition formula for $Z_*^{(1)}$ into a pair of the $Z_*^{(0)}$
models, even before explicit expressions are discussed.

\paragraph{The Selberg integrals.}Explicitly, the Selberg integrals have a form of
the eigenvalue $\beta$-ensembles
\be
Z_{S}^{(0)}\big( u, v, N \big| t \big) = \dfrac{1}{S_0} \int\limits_{0}^{1} dz_1 \ldots dz_{N} \prod\limits_{i < j}^{N} (z_i - z_j)^{2\beta} \prod\limits_{i=1}^{N} z_i^{u} (z_i - 1)^{v} e^{ \beta \sum_{k = 1}^{\infty} t_k z_i^k }
\ee
\be
Z^{(1)}_S\big( A, N \big| q \big) = \dfrac{1}{S_1} \int\limits_{0}^{2\pi} dz_1 \ldots dz_{N} \prod\limits_{i < j}^{N} \theta^{*}\big(z_i - z_j\big|q\big)^{2\beta} \prod\limits_{i=1}^{N} \theta^{*}\big(z_i\big|q\big)^{-\beta N} e^{ I A z_i }
\ee
where $I = \sqrt{-1}$ and $\theta^{*}(z|q) = \sin(z/2) - q \sin(3z/2) + \ldots$ is
the normalized odd theta-function on torus \cite{B1mamo}. The normalization
constants $S_0, S_1$ are needed to satisfy the requirements
$Z_{S}^{(0)}( t = 0 ) = Z^{(1)}_S( q = 0 ) = 1$ implied by the conditions
$B^{(0)}(x = 0) = B^{(1)}(q = 0) = 1$ for properly normalized conformal blocks.
It is these $\beta$-ensembles that we will use to study the pure gauge limit.

\paragraph{Parameters of Selberg integrals.} To study the PGL in terms of the
Selberg integrals, one needs to describe clearly the values of their parameters.
In the spherical case \cite{B0mamo,MMMsurop}, they are given by

\begin{align}
& N_+ = \dfrac{\alpha_1 - \alpha_2 + \alpha - \epsilon_1 - \epsilon_2}{\epsilon_1}, \ \ \ \ u_+ = 2 \dfrac{\alpha_1 - \epsilon_1 - \epsilon_2}{\epsilon_2}, \ \ \ \ v_+ = - 2 \dfrac{\alpha_2}{\epsilon_2}
\label{mAGT1}
 \\
& \nonumber \\
& \ \ \ \ \ \ \ \ \ N_- = \dfrac{\alpha_4 - \alpha_3 - \alpha}{\epsilon_1}, \ \ \ \ \ \ \ u_- = 2 \dfrac{\alpha_4 - \epsilon_1 - \epsilon_2}{\epsilon_2}, \ \ \ \ v_- = - 2 \dfrac{\alpha_3}{\epsilon_2} \label{mAGT2}
\end{align}
\smallskip\\
and, in the toric case \cite{B1mamo}, they are given by

\begin{align}
N = -\dfrac{\alpha_{ext}}{\epsilon_1}, \ \ \ A = \dfrac{2\alpha + \epsilon_1 + \epsilon_2}{\epsilon_2} \equiv \dfrac{2a}{\epsilon_2}, \ \ \ \nu = 3\Delta_{ext} + 3N - 1
\label{mAGT3}
\end{align}
\smallskip\\
where the $\alpha$-parameters are related to the initial $\Delta$-parameters
(conformal dimensions) via
\be
\Delta(\alpha) = \dfrac{\alpha(\epsilon_1+\epsilon_2-\alpha)}{\epsilon_1\epsilon_2}
\ee
and
\be
c = 1 - 6 \left( \dfrac{\epsilon_1+\epsilon_2}{\epsilon_1\epsilon_2} \right)^2
\ee
is the central charge. Here it is convenient to write all these formulas in terms of
the Nekrasov parameters $\epsilon_1,\epsilon_2$, which are
in one-to-one correspondence with the matrix model parameters $\beta$ (the power
of the Vandermonde determinants) and $g_s = g$ (the "string" coupling constant,
aka the genus expansion parameter):
\be
\epsilon_1 = -g\sqrt{\beta}, \ \ \ \epsilon_2 = g/\sqrt{\beta}
\ee
or vice versa
\be
g^2 = - \epsilon_1 \epsilon_2, \ \ \ \beta = -\dfrac{\epsilon_1}{\epsilon_2}
\ee
This completes the list of relations between the parameters, and allows one to look at
the PGL of the Selberg integrals. This limit is simple in terms of the external
dimensions:
\be\label{limit}
\alpha_i\longrightarrow \infty,\ \ \ \ \alpha_i^4x=\hbox{fixed}=\Lambda^4
\ee
$$
\alpha_{ext}\longrightarrow \infty,\ \ \ \ \alpha_{ext}^4q=\hbox{fixed}=\Lambda^4
$$
while in terms of the
matrix model parameters it gets more sophisticated. We will now describe the limit
in terms of $N,u,v$ (for the spherical Selberg models) and of $N,A$
(for the elliptic model).

\paragraph{PGL of Selberg integrals.} Relations (\ref{mAGT1})-(\ref{mAGT2}) imply
that, in the PGL, the parameters $u,v,N$ of the spherical Selberg integrals all tend to infinity.
 However, the same relations indicate that a particular combination of
parameters, that is, $u + v + 2\beta N$ remains finite in the PGL, since it does
not depend on the external dimensions. We find it most convenient to parametrize
this combination by a single variable $n$
\be
{\rm PGL}(u + v + 2\beta N) = \beta n + \beta - 1
\label{PGLPGL}
\ee
which is equal to
\be
n_{\pm} = \pm \dfrac{2a}{\epsilon_1}, \ \ \ a = \alpha - \dfrac{\epsilon_1+\epsilon_2}{2}
\ee
for our "+" and "-" Selberg models. Consequently, the PGL for the spherical Selberg
model looks like a non-trivial double scaling limit, where the parameters $u,v,N$
tend to infinity as
\be
Z_*^{(0)}(n|t) = \lim_{
\stackrel{
\stackrel{u,v,N\rightarrow \infty}{}
}
{u+v+2\beta N \equiv \beta n + \beta - 1}
} Z_{S}^{(0)}\left( u, v, N \ \Big| \ \dfrac{t_k}{(u N + \beta N^2)^k} \right)
\label{PGL0}
\ee
where distinguished combinations $(u N + \beta N^2)_+$ and $(u N + \beta N^2)_-$ in the PGL
are proportional to $\mu_1\mu_2$ and $\mu_3 \mu_4$, respectively ($\mu$'s are mass parameters
of the Nekrasov function, linearly related to $\alpha$'s). This particular rescaling of the
time-variables
$t_k \mapsto t_k (u N + \beta N^2)^{-k}$ is necessary to suppress a growth of
correlators in the model: only with variables defined in this way, the partition
function has a finite PGL. In particular, only with such a rescaling of variables
the decomposition formula (\ref{decoB0}) remains non-trivial in the PGL (\ref{limit})
and, moreover,
turns into formula (\ref{decoB*}), i.e.
\be
Z^{(0)}_{*\pm}(t^{\pm}) = Z_{*}^{(0)}\big( n_{\pm} \big| t^{\pm} \big)
\ee
Similarly, in the toric case, relations (\ref{mAGT3}) imply that the PGL for the
toric Selberg integral is
\be
Z_*^{(1)}(A|\Lambda) = \lim_{
\stackrel{
\stackrel{N\rightarrow \infty, \ q\rightarrow 0}{}
}
{q \beta^2 N^4 \equiv \Lambda^4}
} Z^{(1)}_S\big( A, N \big| q \big)
\label{PGL1}
\ee
where, since no time-variables are introduced, no additional rescalings are required.
As one can see, the PGL's of the Selberg models are quite sophisticated: it is by no
means transparent that eqs.(\ref{PGL0}) or (\ref{PGL1}) do at all have a finite
limit. However, as we shall see below, they do, and the main problem is to give
some constructive description of this limit. This paper is devoted to finding
a (at least, partial) solution to this problem.

\paragraph{PGL of spherical Selberg: BGW model.} As the first (simplest) part of
solution to this problem, in this paper we demonstrate that $Z_{*}^{(0)}$,
the PGL of the Selberg partition function $Z_{S}^{(0)}$ is actually the partition
function of the ($\beta$-deformed) celebrated BGW model \cite{BGW,MMSem,AMMbgw}
of size $n$ and in the character phase \cite{MMSem}:
\be
\boxed{ Z_{*}^{(0)}\big( n \big| t \big) = Z_{BGWc}\Big(n \big| t_k = \tr \Psi^k/k \Big) = \dfrac{1}{{\rm Vol}_\beta(n)} \int\limits_{n \times n} [dU]_{\beta} e^{\beta \big( \tr U^\dagger + \tr {\Psi} U \big)} } = 1 + \dfrac{\beta}{\beta n + 1 - \beta} t_1 + \emph{}
\label{IntClaim}
\ee
$$
\emph{} + \dfrac{\beta^2(\beta n + 2 - 2 \beta)}{(\beta n + 1 - \beta)(\beta n + 2 - \beta)
(\beta n + 1 - 2\beta)} t_1^2 - \dfrac{\beta^2}{(\beta n + 1 - \beta)(\beta n + 2 - \beta)
(\beta n + 1 - 2\beta)} t_2 + \ldots
$$
where the integral over $U$ is the $\beta$-deformed unitary integral, and
${\rm Vol}_{\beta}(n)$ is the $\beta$-deformed volume of the unitary group.
As usual for the BGW model, the time-variables are identified with traces of the
external
field powers $t_k = \tr \Psi^k/k$, and this brings us directly to the topic of
$\beta$-ensembles with external fields, which is somewhat underinvestigated and
not exhaustively covered in the existing (physical) literature. The point is that
the $\beta$-deformations are usually defined for integrals of eigenvalues only.
Whereas the notion of trace (of determinant, etc.) in (\ref{IntClaim}) remains
well-defined as
combinations of the eigenvalues, the treatment of the external field term
$\tr U\Psi$ in (\ref{IntClaim})
deserves some comments. Actually one needs only the $U$-integrals
(averages) of such quantities, they will
be defined in s.\ref{2} with the help of a $\beta$-ensemble version of the
Harish-Chandra-Itzykson-Zuber integral.

In this paper we prove eq. (\ref{IntClaim}) in two independent, but complementary
ways. After the $\beta$-unitary BGW model is defined in s.\ref{2}, in s.\ref{3} we
demonstrate that its Jack expansion (the $\beta$-ensemble counterpart of
the character expansion) coincides with the PGL of the Jack expansion for the
spherical Selberg model. This is just an algebraic exercise, which still may be not
too much transparent. A more conceptual way may be the method of s.\ref{4}, where one
instead takes the PGL of the Virasoro constraints (Ward identities) for the spherical
Selberg model,
and then shows that they coincide with the known Virasoro constraints
\cite{MMSem,AMMbgw} for the
character phase of the BGW $\beta$-ensemble.

Formula (\ref{IntClaim}) fully defines the middle part of
formula (4).
It still remains to explain how
the time derivatives can be taken
in the external field BGW model:
the simplest possibility is
provided by a Fourier-like transform
in the $\beta$-character calculus,
which is reminded in s.\ref{2}.

\paragraph{PGL of elliptic Selberg: double BGW model.} The second, harder part of
the solution, taking the PGL of the elliptic Selberg model is not completely finalized
in the present paper. The corresponding partition function $Z_*^{(1)}(A|\Lambda)$
should be given by some $\beta$-ensemble with the partition function
\be
Z_*^{(1)}(A|\Lambda) = 1 + \frac{2\beta\Lambda^4}{A^2 - (\beta-1)^2}
+ \frac{\beta^2\Lambda^8 \Big(2A^2+\beta - 8(\beta-1)^2\Big)}
{\Big(A^2 - (\beta-1)^2\Big)\Big(A^2 - (2\beta-1)^2\Big)\Big(A^2 - (\beta-2)^2\Big)}
+ \ldots
\label{Elliptic}
\ee
Note that the problem of finding the PGL of the elliptic Selberg model is just the
same as finding the PGL  $B_*(\Delta|\Lambda)$ of the conformal block/Nekrasov
function. Having in mind the general context of the problem and lessons from the
(successful) solution in the spherical case, one can go further in several directions.

One way to solve this problem is to attack it directly from the elliptic side, by
trying to take
the Inozemtsev limit \cite{Ino} of the Ward identities for the elliptic Selberg
model $Z_*^{(1)}$. This remains to be done.

Another way is to make a direct educated guess for what the $\beta$-ensemble
in question should be.
Such an attempt has actually been made long ago in \cite{DVold}.
The conjecture was that $Z_*^{(1)}$ is again a BGW model,
but this time in another phase:
the double-cut DV phase, i.e.
\be
Z_*^{(1)} \stackrel{presumably}{=} Z_{BGW}^{DV_2}
\label{DVconj}
\ee
This suggestion can be checked, for example, by a derivation
of AMM decomposition of $Z_{BGW}^{DV_2}$ and comparison with
the second equality in (\ref{decoB*}). This also remains to be done.

The third possibility is to apply the character
calculus of \cite{Unitint}, either to the elliptic
Selberg integral itself or to the decomposition
formula (\ref{decoB*}). This is what we do in the present paper, in s.\ref{5} below.
Applying the character calculus to the decomposition formula (\ref{decoB*}),
one obtains a double $\beta$-unitary-ensemble
{\fontsize{9.5pt}{0pt}
\be
\boxed{
Z_*^{(1)}(A|\Lambda) =
\int \dfrac{[dU]_{\beta}}{{\rm Vol}_{\beta}\left( n_+ \right)} \int \dfrac{[d{\tilde U}]_{\beta}}{{\rm Vol}_{\beta}\left( n_- \right)} \ Z_{BGWc} \left( m_+ \Big| s_k \right) \ Z_{BGWc} \left( m_- \Big| \tilde s_k \right)
\det\Big( 1 - \Lambda^4 U^{+} \otimes {\tilde U}^{+} \Big)^{2\beta}
}
\label{2Matrix}
\ee}

\noindent
where $k s_k = \tr U^k, k {\tilde s}_k = \tr {\tilde U}^k$ and
$m_{\pm} = n_{\pm} + (\beta-1)/\beta$.
Thus, $Z_{BGW}$'s in the integrand are actually
functions of $U$ and $\tilde U$, while their
conjugates $U^\dagger$ and $\tilde U^\dagger$
enter through the mixing (intertwining)
determinant.
Note that the $\beta$-unitary integrals in
(\ref{2Matrix}) have sizes $n_{\pm} = \pm 2a/\epsilon_1$, while the BGW models in
the integrand have sizes $m_{\pm} = \pm 2\alpha/\epsilon_1$. As explained in
s.\ref{5}, this shift of sizes can be seen as a natural property of the Fourier
transformation for the $\beta$-ensembles.

Eq.(\ref{2Matrix}) or, at least, the
first terms of its $\Lambda$-expansion
(\ref{Elliptic}) can be checked in
practice by substituting (\ref{IntClaim}) into (\ref{2Matrix}) and taking the
remaining averages
\be
\int\limits_{n \times n} f(U) [dU]_{\beta} \equiv \int\limits_{-\pi}^{\pi} d\phi_1 \ldots \int\limits_{-\pi}^{\pi} d\phi_n f\left(e^{I \phi_1}, \ldots, e^{I \phi_n}\right) \prod\limits_{a < b}^{n} \left| e^{I \phi_a} - e^{I \phi_b} \right|^{2\beta}
\ee
which is valid when $f(U)$ is an invariant function (depends only on eigenvalues or,
what is the same, on traces of powers of $U$). The averages, necessary to reproduce the
first two orders in $\Lambda$, are
{\fontsize{10pt}{0pt}
\be
\int\limits_{n \times n} \dfrac{[dU]_{\beta}}{{\rm Vol}_{\beta}\left( n \right)} \tr U \tr (U^{+}) = \dfrac{n}{\beta n + \beta - 1}, \ \ \ \int\limits_{n \times n} \dfrac{[dU]_{\beta}}{{\rm Vol}_{\beta}\left( n \right)} \tr U^2 \tr (U^{+})^2  = \dfrac{2n(\beta^2 n^2-\beta^2+\beta-1)}{(\beta n+\beta)(\beta n-1)(\beta n+\beta-1)}
\ee
\be
\int\limits_{n \times n} \dfrac{[dU]_{\beta}}{{\rm Vol}_{\beta}\left( n \right)} \tr U \tr U \tr (U^{+})^2 = \int\limits_{n \times n} \dfrac{[dU]_{\beta}}{{\rm Vol}_{\beta}\left( n \right)} \tr U^2 \tr (U^{+}) \tr (U^{+}) = \dfrac{2n(\beta-1)}{(\beta n+\beta)(\beta n-1)(\beta n+\beta-1)}
\ee
\be
\int\limits_{n \times n} \dfrac{[dU]_{\beta}}{{\rm Vol}_{\beta}\left( n \right)} (\tr U)^2 (\tr U^{+})^2 = \dfrac{2n(\beta n^2-1)}{(\beta n+\beta)(\beta n-1)(\beta n+\beta-1)}
\ee}
Using them, one easily verifies that (\ref{2Matrix}) reproduces (\ref{Elliptic}),
so that operationally the model is well-defined and gives correct results. However,
there are many conceptual questions left, in particular, the relation of the model
(\ref{2Matrix}) to the Inozemtsev limit of the elliptic Selberg model and,
most importantly, to the double-cut BGW integrals. If the DV conjecture
(\ref{DVconj}) is true, the model (\ref{2Matrix}) should probably be interpreted as
an integral representation of $Z_{BGW}^{DV_2}$. We do not consider this topic in the
present paper, this will be done elsewhere.

\section{Comments on the definition of $\beta$-deformed BGW model\label{2}}

Generalization of unitary integrals to $\beta\neq 1$ deserves comments.
The problem is to give some concrete definition for the $\beta$-deformed unitary
integral of the form
\be
\int\limits_{n \times n} f(U) [dU]_{\beta}
\ee
which for $\beta = 1$ is the well-defined integral over the compact Lie group $U(n)$
with the Haar invariant measure $[dU]_{\beta=1}$. We are unaware of any similar
group theory definition for \emph{generic} $\beta$ (occasionally, such a definition
exists for $\beta = 1/2$ and $\beta = 2$ in terms of the groups $O(n)$ and $Sp(n)$,
respectively). Instead, various other, more or less natural definitions can be
suggested.

The simplest case is when $f(U)$ is an invariant function, i.e. it depends only on
the traces of powers of $U$ or, equivalently, only on the eigenvalues of $U$. In this
case, the most natural definition of $\beta$-generalization, motivated by
consideration of the above-mentioned three group integrals ($U(n)$ for $\beta = 1$,
$O(n)$ for $\beta = 2$ and $Sp(n)$ for $\beta = 1/2$) is the following eigenvalue
integral:
\be
\int\limits_{n \times n} f(U) [dU]_{\beta} \equiv \int\limits_{-\pi}^{\pi} d\phi_1 \ldots \int\limits_{-\pi}^{\pi} d\phi_n \ f\left({\rm diag}\left[ e^{I \phi_1}, \ldots, e^{I \phi_n}\right] \right) \prod\limits_{a < b}^{n} \left| e^{I \phi_a} - e^{I \phi_b} \right|^{2\beta}
\label{betaunitary}
\ee
i.e. the role of the $\beta$-deformation is just to have the power $2\beta$
of the Van-der-monde determinant. It is this definition which is most commonly
recalled when the words "$\beta$-ensemble" are mentioned.

However, for the purpose of present paper this definition \emph{is not enough}.
The case when the integrand $f(U)$ depends only
on the eigenvalues, does {\it not} cover {\it all}
the {\it eigenvalue models} \cite{GKM,UFN3}.
In particular, the main object of the present paper, which we use to describe the
PGL of the Selberg integrals, has an integrand which is \emph{not} a function of
the eigenvalues of $U$ only, it involves an "external field" matrix $\Psi$ in the
following way
\be
Z_{BGW}(\Psi) = \int\limits_{n \times n} \dfrac{[dU]_{\beta}}{{\rm Vol}_{\beta}\left( n \right)} e^{\beta \big( \tr U^\dagger + \tr {\Psi} U \big)}
\ee
so that definition (\ref{betaunitary}) is not applicable. Note that for $\beta = 1$
the integral $Z_{BGW}(\Psi)$ obviously depends only on the eigenvalues of the
matrix $\Psi$. We can quite naturally assume the same property to hold for all
$\beta$. After that, the two options still remain: to consider the integral as a
function of traces of \emph{positive} powers $k t_k = \tr \Psi^k$ or of
\emph{negative} powers $k \tau_k = \tr \Psi^{-k}$. As is well-known in the theory of
the ordinary ($\beta = 1$) BGW model, these two choices lead to different results,
commonly known as the \emph{character phase} $Z_{BGWc}(t)$ and the
\emph{Kontsevich phase} $Z_{BGWk}(\tau)$, respectively \cite{MMSem}.
In this paper we are interested in the definition of the first one, $Z_{BGWc}(t)$.

Our definition refers to the Harish-Chandra-Itzykson-Zuber (HCIZ) integral over
the unitary matrix $V$,
$
Z_{IZ}(t,s) =
\int\limits_{n \times n} \dfrac{[dV]_{\beta}}{{\rm Vol}_{\beta}\left( n \right)} e^{\beta \tr {\Psi}VUV^\dagger}
$
which is a function of $kt_k = \tr {\Psi}^k$
and $ks_k = \tr U^k$.
To see this at $\beta=1$, it is enough to diagonalize the
matrices, ${\Psi} = V_{\Psi}{\Psi}_dV_{\Psi}^\dagger$ and
$U = V_U U_d V_U^\dagger$ and use invariance
of the Haar measure $[dV]$ to change $V \longrightarrow V_{\Psi}^\dagger V V_U$. Then
\be
Z_{IZ}(t,s)\Big|_{\beta=1} = \int\limits_{n \times n} \dfrac{[dV]}{{\rm Vol}\left( n \right)} e^{\tr {\Psi}VUV^\dagger} =
\int\limits_{n \times n} \dfrac{[dV]}{{\rm Vol}_{\beta}\left( n \right)} e^{\sum_{i,j} {\Psi}_iU_j |V_{ij}|^2 }
\label{IZ1}
\ee
is indeed a function of the eigenvalues $\{{\Psi}_i\}$ and $\{U_j\}$
only, and thus of $t$ and $s$
(the symmetry under permutation of the eigenvalues
is obvious).
Once $Z_{IZ}$ is defined, one can write
\be
Z_{BGWc}(t) = \int\limits_{n \times n} \dfrac{[dU]_{\beta}}{{\rm Vol}_{\beta}\left( n \right)} Z_{IZ}(t,s) e^{\beta \tr U^\dagger}
\label{BGWIZ}
\ee
which is now a function of the positive time-variables $t$.
Moreover, the integrand also depends only on the
eigenvalues of the integration matrix-variable $U$, so that
definition (\ref{betaunitary}) is applicable.

Thus, to define $Z_{BGWc}(t)$, it suffices to define the
Itzykson-Zuber integral (\ref{IZ1}) in some independent way. Such a
possibility is provided by the character calculus \cite{Unitint}:
for $\beta=1$, the IZ integral can be defined as an expansion \be
Z_{IZ}(t,s) = \sum_R \frac{d_R}{D_R(n)} \chi_R(t)\chi_R(s) , \ \ \ \
\beta=1 \ee where $\chi_R$ are the characters, and $d_R,D_R$ are their values at particular
points $d_R = \chi_{R}(t_k = \delta_{k,1})$, $D_R(n) = \chi_{R}(t_k =
n/k)$. The quantities $d_R$ and $D_R$ have an important representation theory meaning: they are
dimensions of representations of the symmetric- and $GL(n)$- groups, respectively
(labeled by the Young diagram $R$). Thus, for arbitrary $\beta$, one can naturally {\it define}
\be Z_{IZ}(t,s) = \sum_R
\frac{d_R}{D_R(n)}
j_R(t) j_R(s), \ \ \ \ \ \forall \beta \label{IZdef} \ee where $j_R$
are the well-known $\beta$-characters (actually, the properly
normalized Jack polynomials \cite{Sh1}, reviewed in detail in
Appendix 1 of the present paper) and $d_R,D_R$ are their values at particular points
$d_R = j_{R}(t_k = \delta_{k,1})$, $D_R(n) = j_{R}(t_k = n/k)$. For arbitrary $\beta$, these
quantities do not have any straightforward representation theory meaning, they can be only
thought of as $\beta$-deformations of the dimensions of symmetric and general linear groups.
Note that the
$n$-dependence in formula (\ref{IZdef}) emerges only due to the $n$-dependent quantity $D_R(n)$.

Eqs. (\ref{BGWIZ}) and (\ref{IZdef}) provide our {\it constructive} definition of the BGW
partition function in the character phase. It should be emphasized,
that \textbf{in this way the $\beta$-deformation of any unitary eigenvalue model with no more
than one external field in character phase can be defined}. This is
enough for our purposes here, but not enough in principle: to study
the $\beta$-deformations of the Kontsevich phases or the $\beta$-deformations of the
models
with multiple external fields (known as non-eigenvalue models), some other definitions
have to be invented. Hopefully, there exists a unifying framework for
the $\beta$-deformations of
\emph{any} matrix model, not necessarily of the eigenvalue type. This framework
remains to be discovered.

Note that definition (\ref{IZdef}) already appeared in mathematical
literature, see, for example, \cite{Sh2}. It goes without saying
that it reproduces the HCIZ integrals not only for unitary
($\beta=1$), but also for orthogonal ($\beta = 1/2$) and symplectic
($\beta=2$) matrices.

Given such a definition of $Z_{BGWc}$ $\beta$-ensemble, it is possible to study
if it indeed coincides with the PGL $Z_{*}^{(0)}$ of the spherical Selberg
$\beta$-ensemble. The next two sections are devoted to two ways of proving this.
The first method makes a direct use of properties of the $\beta$-characters $j_R$,
while the second method relies upon Ward identities for the $\beta$-ensembles.

\section{Eq. (\ref{IntClaim}) via Jack expansion\label{3}}

To check the equality $Z^{(0)}_* = Z_{BGWc}$, we expand the both quantities in the basis of
Jack polynomials.

\paragraph{The right hand side.}For $Z_{BGWc}$, eqs.(\ref{BGWIZ}) and (\ref{IZdef})
imply that
\be
Z_{BGWc}(t) = \int Z_{IZ}(t,s) e^{\beta \tr U^\dagger} [dU]_\beta =
\sum_R j_R(t) \frac{d_R}{D_R(n)} \int j_R(s) e^{\beta \tr U^\dagger} [dU]_\beta
\ee
To deal with the r.h.s. integral, one needs just two properties of
the $\beta$-characters: the completeness
\be
\exp\left( \beta\sum_k k t_k {\tilde t}_k \right) = \sum\limits_R j_R(t) j_R({\tilde t})
\label{Complete}
\ee
and the Haar orthogonality
\be\label{Hort}
\int j_R(s) j_{{\tilde R}}(s^{-1}) [dU]_\beta = \delta_{R,{\tilde R}}
\dfrac{ D_R(n) }{D_R(n-\delta)}, \ \ \ \delta = \dfrac{\beta-1}{\beta}
\ee
where $n$ is the size of $U$. The first identity (with $k t_k = \tr (U^\dagger)^k,
k {\tilde t}_k = \delta_{k,1}$) implies that
\be
e^{\beta \tr U^\dagger} = \sum\limits_{R} d_R j_R\big(s^{-1}\big)
\ee
and the second property allows one to perform the integration over $U$:
\begin{align}
Z_{BGWc}(t) = \int Z_{IZ}(t,s) e^{\beta \tr U^\dagger} [dU]_\beta & =
\sum\limits_R \sum\limits_{{\tilde R}} j_R(t) \frac{d_R d_{\tilde R}}
{D_R(n)} \int j_R(s) j_{{\tilde R}}(s^{-1}) [dU]_\beta \\ \nonumber & \\ &  =
\sum_R \frac{d_R^2}{D_R(n-\delta)} j_R(t)
\end{align}
The expansion obtained
\be
\boxed{
Z_{BGWc}(t) = \sum_R \frac{d_R^2}{D_R(n-\delta)} j_R(t)
\label{BGWexpan}
}
\ee
is of course a $\beta$-deformation of the well-known
character expansion \cite{Unitint}
\be
Z_{BGWc}(t) = \sum_R \frac{d_R^2}{D_R(n)} \chi_R(t), \ \ \ \ \ \ \beta=1
\ee
but not quite a naive one (because of the non-trivial $\delta$-shift in the denominator).
Note that the matrix sizes enter the character expansions only through the explicit
factors $D_R$. Let us now derive a similar expansion for the Selberg partition function.

\paragraph{The left hand side.}The Jack expansion for the Selberg $\beta$-ensemble
\be
Z_{S}^{(0)}\big( u, v, N \big| t \big) = \dfrac{1}{S_0} \int\limits_{0}^{1} dz_1 \ldots dz_{N} \prod\limits_{i < j}^{N} (z_i - z_j)^{2\beta} \prod\limits_{i=1}^{N} z_i^{u} (z_i - 1)^{v} e^{ \beta \sum_{k = 1}^{\infty} t_k z_i^k }
\ee
using the same completeness condition (\ref{Complete}) can be written as
\be
Z_{S}^{(0)}\big( u, v, N \big| t \big) = \sum\limits_R \Big< j_R \Big> j_R(t)
\ee
where the coefficients $\Big< j_R \Big>$ are averages of the Jack polynomials of the
$z$-variables,
\be
\Big< j_R \Big> = \dfrac{1}{S_0} \int\limits_{0}^{1} dz_1 \ldots dz_{N} \ j_R\left( t_k = \sum_i z_i^k/k \right) \ \prod\limits_{i < j}^{N} (z_i - z_j)^{2\beta} \prod\limits_{i=1}^{N} z_i^{u} (z_i - 1)^{v}
\ee
Fortunately, the averages of Jack polynomials in the Selberg model are well-known to
be simple quantities \cite{ItoJack}. They factorize nicely, and can be generally expressed
by the Kadell formula
\be
\Big< j_{R} \Big> = \dfrac{d_R}{\beta^{|R|}} \ \dfrac{[N \beta]_R [u + (N-1)\beta + 1]_R}{[u+v+2N\beta+2-2\beta]_R}
\label{JackAverage}
\ee
where notation $[...]_R$ (see Appendix 1) stands for
\begin{align}
[x]_R = \beta^{|R|} \cdot \dfrac{D_R(x)}{d_R} = \prod\limits_{(i,j) \in R} \Big( x - \beta ( i - 1) + (j - 1) \Big)
\end{align}
\smallskip\\
Using this Kadell formula, it is immediate to take the PGL (\ref{PGLPGL}):
\be
\Big< j_{R} \Big>_* = \dfrac{d_R^2}{D_R(n - \delta)} (u N + \beta N^2)^{|R|}, \ \ \ \delta = \dfrac{\beta-1}{\beta}
\label{JackAveragePGL}
\ee
As one can see, the correlators grow like $(u N + \beta N^2)^{|R|}$ in the PGL.
This is actually the reason to introduce the corresponding rescaling of time-variables
in the PGL of the partition function:
\be
Z_{*}^{(0)}\big( n \big| t \big) = \lim_{
\stackrel{
\stackrel{u,v,N\rightarrow \infty}{}
}
{u+v+2\beta N \equiv \beta n + \beta - 1}
} Z_{S}^{(0)}\left( u, v, N \ \Big| \ \dfrac{t_k}{(u N + \beta N^2)^k} \right) = \sum\limits_R  \Big< j_R \Big>_* \dfrac{ j_R(t) }{(u N + \beta N^2)^{|R|}}
\ee
Substituting here eq.(\ref{JackAveragePGL}), one finds
\be
\boxed{
Z_{*}^{(0)}\big( n \big| t \big) = \sum_R \dfrac{d_R^2}{D_R(n - \delta)} j_R(t)
\label{PGLexpan}
}
\ee
what precisely coincides with (\ref{BGWexpan}). In this way, the PGL limit of the
Kadell formula reproduces the BGW model: we conclude that $Z^{(0)}_* = Z_{BGWc}$,
since their Jack expansions are just the same. Let us now pass to the second method
of proving this statement.

\section{Eq. (\ref{IntClaim}) via Virasoro constraints\label{4}}

Another way to see that $Z^{(0)}_* = Z_{BGW}$ is to study the PGL of the Virasoro
constraints for the Selberg model. Just as any matrix model (or the $\beta$-ensemble),
the Selberg model can be characterized by certain linear differential equations, which
arise as a consequence of the reprarametrization invariance of the multiple integral
(i.e. as the Ward identities). In the case of the Selberg model, these linear
differential equations have the form
\begin{align*}
\left[ \Big( u + v + 2 \beta N + (k + 1)(1 - \beta) \Big) \dfrac{\partial}{\partial t_k} + \beta \sum\limits_{m} m t_{m} \dfrac{\partial}{\partial t_{k + m}} + \sum\limits_{a + b = k} \dfrac{\partial^2}{\partial t_a \partial t_b} + v \sum\limits_{h = 1}^{k-1} \dfrac{\partial}{\partial t_h} \right] Z_S(t_1, t_2, \ldots ) =
\end{align*}
\begin{align}
= \beta \left( N + \sum\limits_{i = 1}^{m-1} k_i + u N + N(N-1)\beta \right) Z_S(t_1, t_2, \ldots ), \ \ \ k > 0
\end{align}
Their derivation is given in Appendix 2. These equations completely determine the
partition function of the Selberg model, therefore, one can take the PGL in these
equations, not in the integral. In the PGL, all the three parameters $u,v,N$ of the
Selberg models tend to infinity in such a way that their combination
$u + v + 2 \beta N$ is held finite.
In this limit, the Virasoro constraints get simplified (many terms can be thrown out)
and turn into
\be
\left[ \Big( u + v + 2 \beta N - (k + 1)(\beta - 1) \Big) \dfrac{\partial}{\partial t_k} + \beta \sum\limits_{m} m t_{m} \dfrac{\partial}{\partial t_{k + m}} + \sum\limits_{a = 1}^{k-1} \dfrac{\partial^2}{\partial t_a \partial t_{k-a}} + \beta \delta_{k, 1} \right] Z_*^{(0)}(t) = 0
\label{AsymptVirasoro}
\ee
for any $k > 0$, and these are precisely the Ward identities for integral (\ref{BGWIZ}). They can be
found in \cite{MMSem,AMMbgw}, of course, only for the most popular case of
$\beta = 1$. Despite there are no doubts that eqs. (\ref{AsymptVirasoro}) hold for
arbitrary $\beta$, in principle, it would be nice to derive them directly from the
definition of the BGW $\beta$-ensemble (in the present paper, the role of such a
definition is played by the character calculus). This remains to be done.

\section{Eq. (\ref{2Matrix}) via Fourier transform\label{5}}

Definition (\ref{BGWexpan}) can be directly used to convert
the action of the intertwining operator in (\ref{decoB*}) into
a kind of a 2-matrix BGW model.
This can be done via the Fourier transform,
widely used in the character calculus \cite{Unitint}.
The basic and most important Fourier relation has the form
\be
j_R(t) = \dfrac{D_R(n - \delta)}{D_R(n)} \int\limits_{n \times n} \dfrac{[dU]_{\beta} }{{\rm Vol}_{\beta}(n)} j_R(s) \exp\left( \beta \sum_k k t_k s_k \right)
\ee
where $k s_k = \tr U^k$. This relation is a direct consequence of the completeness
condition (\ref{Complete}) and the Haar orthogonality (\ref{Hort}): one substitutes
(\ref{Complete}) into the r.h.s. and calculates
integral with the help of (\ref{Hort}).
Contracting both sides of this relation with the coefficients of the BGW expansion, one
finds
\be
Z_{BGWc}(n|t) = \sum_R \frac{d_R^2}{D_R(n-\delta)} j_R(t) = \int\limits_{n \times n} \dfrac{[dU]_{\beta} }{{\rm Vol}_{\beta}(n)} \sum_R \frac{d_R^2}{D_R(n)} j_R(s) \exp\left( \beta \sum_k k t_k s_k \right)
\ee
and the sum in the r.h.s. is the BGW partition function again, only of a different size:
\be
Z_{BGWc}(n|t) = \int\limits_{n \times n} \dfrac{[dU]_{\beta} }{{\rm Vol}_{\beta}(n)} Z_{BGWc}\left(n + \delta \Big|s\right) \exp\left( \beta \sum_k k t_k s_k \right)
\ee
This relation can be understood as a Fourier transform for the BGW $\beta$-ensemble.
It is important for two reasons.

First, it allows one to write $t$-derivatives of the partition function (correlators)
as integrals: say,
\be
\frac{\p^m}{\p t_{k_1}\ldots\p t_{k_m}} Z_{BGWc}(n|t) \Big|_{t = 0} =
\int\limits_{n \times n} \dfrac{[dU]_{\beta} }{{\rm Vol}_{\beta}(n)}
Z_{BGWc}\left(n + \delta \Big|s\right) \big( \beta \tr U^{k_1} \big) \ldots \big( \beta \tr U^{k_m} \big)
\ee
whereas without the Fourier transform one could only write
\be
\frac{\p^m}{\p t_{k_1}\ldots\p t_{k_m}} Z_{BGWc}(n|t) \Big|_{t = 0} =
\frac{\p^m}{\p t_{k_1}\ldots\p t_{k_m}}
\int\limits_{n \times n} \dfrac{[dU]_{\beta} }{{\rm Vol}_{\beta}(n)} e^{\beta \big( \tr U^{+} + \tr U \Psi \big)} \Big|_{\Psi = 0}
\ee
which is largely a symbolical notion: the integrand in the r.h.s. does not depend on
the $t$-variables, only the integral does. Thus, the Fourier transform appears to be a
convenient tool in this case.

Second, the Fourier transform allows one to convert the decomposition operator into an
integral. Indeed, we have
\be
Z_{*}^{(1)} = \exp\left( -\dfrac{2}{\beta} \sum\limits_{k = 1}^{\infty} \dfrac{\Lambda^{4k}}{k} \dfrac{\partial^2}{\partial t^{+}_k \partial t^{-}_k} \right) Z_{BGWc}\big( n_+ \big| t^{+}\big) Z_{BFWc}\big( n_- \big| t^{-} \big)\Big|_{t=0} =
\ee
Substituting each of the BGW partition functions by its Fourier transform, one then
finds
\be\label{56}
Z_{*}^{(1)} = \int \dfrac{[dU]_{\beta}}{{\rm Vol}_{\beta}\left( n_+ \right)} \int \dfrac{[d{\tilde U}]_{\beta}}{{\rm Vol}_{\beta}\left( n_- \right)} \ Z_{BGWc} \left( m_+ \Big| s_k \right) \ Z_{BGWc} \left( m_- \Big| \tilde s_k \right)
\det\Big( 1 - \Lambda^4 U^{+} \otimes {\tilde U}^{+} \Big)^{2\beta}
\ee
where $m_{\pm} = n_{\pm} + (\beta-1)/\beta$ are the shifted sizes. Such a shift looks
like a natural property of the Fourier transforms for the $\beta$-ensembles. For
$\beta = 1$, the shift vanishes.

\section{Conclusion}

In this paper, we have obtained two BGW-model-representations of the Selberg models
($\beta$-ensembles), elementary constitutients of the conformal blocks, in the pure
gauge limit. The spherical Selberg model after the PGL turned out to become the
character phase of the BGW model, while the elliptic Selberg model after the PGL is
converted into some other model $X$. In this paper we succeeded in rewriting $X$ as a
double BGW model, but we still expect some simpler representations for this model.
A possible candidate on the role of such a simpler representation is the double-cut
BGW model, suggested long ago by Dijkgraaf and Vafa \cite{DVold}. If this is correct,
then the double BGW model of the present paper is yet another integral representation
for this double-cut model.

\section*{Acknowledgements}

We are grateful to H.Itoyama, H.Kanno, T.Oota and other
participants of the Japanese-Russian JSPS/RFBR workshop in Moscow
(September, 2010) for the discussions. We also thank A.Alexandrov for a useful discussion.
Our work is partly supported by Ministry of Education and Science of
the Russian Federation under contract 02.740.11.0608 , by RFBR
grants 10-02-00509-  (A.Mir.), and 10-02-00499 (A.Mor. \& Sh.Sh.), by joint grants
09-02-90493-Ukr, 09-01-92440-CE, 09-02-91005-ANF, 10-02-92109-Yaf-a. The work of A.Morozov
was also supported in part by CNRS.

\pagebreak

\section*{Appendix 1. Properties of Jack polynomials}

The Jack polynomials form an important class of symmetric polynomials, which are often
useful in calculations
with arbitrary $\beta$-ensembles, not only of the Dotsenko-Fateev type. For
convenience, in this Appendix we
list several basic formulas related to the Jack polynomials, which are well-known but
scattered in the literature.

\subsection*{Symmetric polynomials}
Symmetric polynomials $f(z_1, \ldots, z_N)$ of
given degree $\deg f$
form a linear space of finite dimension, with the basis vectors labeled by the Young
diagrams $Y = Y_1 \geq Y_2 \geq \ldots$.
Frequently used bases are the following: the power sums $s_Y = \prod_i s_{Y_i}$, where
$s_k(z) = \sum_i z_i^k$ are
the Newton power sums; the elementary symmetric polynomials $e_Y = \prod_i e_{Y_i}$,
where $e_k = $ the coefficient of
$x^k$ in $\prod_i (1 + x z_i)$; the monomial functions $m_Y = $ symmetrization of
$\prod_i z_i^{Y_i}$. The transition
matrices between these bases have the form

\begin{align*}
e_1 = m_1, \ \ e_{11} = m_2 + 2 m_{11}, \ \ e_{2} = m_{11}, \ \ e_{111} = m_3 + 3 m_{21} + 6m_{111}, \ \ e_{21} = m_{21} + 3m_{111}, \ \ e_{3} = m_{111}, \ \ \ldots
\end{align*}
\begin{align*}
m_1 = s_1, \ \ m_{11} = s_{11}/2 - s_2/2, \ \ m_{2} = s_{2}, \ \ m_{111} = s_{111}/6 - s_{21}/2 + s_3/3, \ \ m_{21} = s_{21} - s_3, \ \ m_{3} = s_{3}, \ \ \ldots
\end{align*}
\begin{align*}
s_1 = e_1, \ \ s_{11} = e_{11}, \ \ s_{2} = e_{11} - 2 e_2, \ \ s_{111} = e_{111}, \ \ s_{21} = e_{111} - 2 e_{21}, \ \ s_{3} = e_{111} - 3 e_{21} + 3 e_{3}, \ \ \ldots
\end{align*}
\smallskip\\
Often, instead of power sums $s_k(z) = \sum_i z_i^k$ the time-variables $t_k = s_k/k$ are used.

\subsection*{Jack polynomials}

The Jack polynomials $J_Y$ are the polynomial eigenfunctions

\begin{align}
{\hat W} J_{Y} = \sum_i Y_i \big( Y_i - 2 \beta i - 1 \big) J_Y
\end{align}
\smallskip\\
of the ${\hat W}$-like operator

\begin{align}
{\hat W} = \sum\limits_{k = 1}^{\infty} \Big( k(k-1) - \beta k(k+1) \Big) s_k \dfrac{\partial}{\partial s_k} + \sum\limits_{k,m = 1}^{\infty} \left( km s_{k+m} \dfrac{\partial^2}{\partial s_k \partial s_m} + \beta (k + m) s_k s_m \dfrac{\partial}{\partial s_{k+m}} \right)
\end{align}
\smallskip\\
normalized with a condition $J_Y = m_Y + \ldots$ (i.e. the coefficient in front of $m_Y$ is
equal to unity). Explicitly, several first Jack polynomials with this normalization convention
have the form

\begin{align*}
& J_{1}(s_k) = s_{1}\\&
\\&
J_{2}(s_k) =  \dfrac{s_2 + \beta s_{11}}{\beta + 1}, \ \ \ \ J_{11}(s_k) =  \dfrac{1}{2} \big( s_{1}^2 - s_2 \big) \\&
\\&
J_{3}(s_k) = \dfrac{2 s_3 + 3 \beta s_{1} s_2 + \beta^2 s_{1}^3}{(\beta + 1)(\beta + 2)}, \ \ \ \  J_{21}(s_k) = \dfrac{(1-\beta) s_{1} s_2 - s_3 + \beta s_{1}^3 }{(\beta + 1)(\beta + 2)}, \ \ \ \
J_{111}(s_k) = \dfrac{1}{6} s_{1}^3 - \dfrac{1}{2} s_{1} s_2 + \dfrac{1}{3} s_3 \\
\end{align*}

\subsection*{Normalized Jack polynomials: $\beta$-characters}

Jack polynomials are orthogonal with respect to the following scalar product:

\begin{align}
\Big< J_A \Big| J_B \Big> = \delta_{AB} ||J_A||^2, \ \ \ \ \Big< s_A \Big| s_B \Big> \equiv \left. \prod_j \dfrac{B_j}{\beta} \dfrac{\partial}{\partial s_{B_j}} \prod_i s_{A_i} \right|_{p = 0}
\end{align}
\smallskip\\
which is often called \emph{intersection product}. The expression for the norm is rather
complicated:

\begin{align}
||J_A||^2 = \dfrac{Q_Y}{P_Y}
\end{align}
\smallskip\\
where $P_Y$ and $Q_Y$ are the two "Young normalization factors"

\begin{align}
P_Y = \prod\limits_{(i,j) \in Y} \Big( \beta ( Y^{\prime}_j - i) + (Y_i - j) + \beta \Big)
\end{align}
\begin{align}
Q_Y = \prod\limits_{(i,j) \in Y} \Big( \beta ( Y^{\prime}_j - i) + (Y_i - j) + 1 \Big)
\end{align}
\smallskip\\
which have a form of products over the cells $(i,j)$ of the Young diagrams.
In applications of the Jack polynomials to matrix models, it is often convenient to normalize
them with respect to their norm:

\begin{align}
j_Y = \dfrac{J_Y}{||J_Y||}
\end{align}
\smallskip\\
Such normalized polynomials $j_Y$ satisfy

\begin{align}
\Big< j_A \Big| j_B \Big> = \delta_{AB}
\end{align}
\smallskip\\
and, in general, formulas in terms of $j_Y$ are simpler than formulas in terms of $J_Y$.
However, the polynomials $j_Y$ themselves are more complicated and depend on
$\beta$ irrationally. Explicitly, several first Jack polynomials with this normalization
convention have the form

\begin{align*}
& j_{1}(s_k) = \sqrt{\beta} s_{1}, \ \ \ j_{2}(s_k) =  \sqrt{\dfrac{\beta(\beta+1)}{2}} \dfrac{s_2 + \beta s_{1}^2}{\beta + 1}, \ \ \ \ j_{11}(s_k) = \sqrt{\dfrac{\beta^2}{2\beta+2}} \Big( s_{1}^2 - s_2 \Big) \\&
\\&
j_{3}(s_k) = \sqrt{\dfrac{\beta(\beta+1)(\beta+2)}{6}} \dfrac{2 s_3 + 3 \beta s_{1} s_2 + \beta^2 s_{1}^3}{(\beta + 1)(\beta + 2)}, \ \ \ \  j_{21}(s_k) = \sqrt{\dfrac{(2\beta+1) \beta^2}{\beta+2}} \dfrac{(1-\beta) s_{1} s_2 - s_3 + \beta s_{111} }{(\beta + 1)(\beta + 2)}
\\&
\\&
j_{111}(s_k) = \sqrt{\dfrac{6\beta^3}{(2\beta+1)(\beta+1)}} \Big( \dfrac{1}{6} s_{1}^3 - \dfrac{1}{2} s_{1} s_2 + \dfrac{1}{3} s_3 \Big) \\
\end{align*}

\subsection*{Particular values of Jack polynomials $d_R$ and $D_R$}

Frequently encountered quantities are values of the Jack polynomials at particular points
$s_k = \delta_{k,1} = (1,0,0,\ldots)_k$ and $s_k = n$ (equivalently, $t_k = \delta_{k,1}$ and
$t_k = n/k$). The Jack polynomials have the following values at these points:

\begin{align}
J_Y (s_k = \delta_{k,1}) = \dfrac{\beta^{|Y|}}{P_Y}
\end{align}
\begin{align}
J_Y (s_k = n) = \dfrac{[\beta n]_Y}{P_Y}
\end{align}
\smallskip\\
where $[...]_Y$ is the so-called "Young-Pochhammer symbol"

\begin{align}
[x]_Y = \prod\limits_{(i,j) \in Y} \Big( x - \beta ( i - 1) + (j - 1) \Big)
\end{align}
\smallskip\\
where $Y$ is any Young diagram and $Y^{\prime}$ stands for its transposed diagram. Note that
the Young-Pochhammer symbol is a generalization of the classical Pochhammer falling and rising
factorials

\begin{align}
(x)_n = x(x-1)\ldots(x-n+1) = \dfrac{\Gamma(x+1)}{\Gamma(x-n+1)}
\end{align}
\begin{align}
(x)^{(n)} = x(x+1)\ldots(x+n-1) = \dfrac{\Gamma(x+n)}{\Gamma(x)}
\end{align}
\smallskip\\
which simply correspond to the row and coloumn diagrams $Y$, respectively. For the
normalized Jack polynomials $j_Y$, the values under consideration are usually denoted
through $d_Y$ and $D_Y$, respectively:

\begin{align}
d_Y = j_Y (s_k = \delta_{k,1}) = \dfrac{\beta^{|Y|}}{\sqrt{P_Y Q_Y}}
\end{align}
\begin{align}
D_Y(n) = J_Y (s_k = n) = \dfrac{[\beta n]_Y}{\sqrt{P_Y Q_Y}}
\end{align}
\smallskip\\
This implies, that

\begin{align}
D_Y(n) = d_Y \dfrac{[\beta n]_Y}{\beta^{|Y|}}
\end{align}
\smallskip\\
which is useful to estimate the large-$n$ behaviour of $D_Y(n)$.

\subsection*{Relation to $\beta$-deformed unitary integrals.}

The Jack polynomials are also orthogonal with respect to the integral product

\begin{align}
\Big< s_A \Big| s_B \Big>_{n} \equiv \dfrac{ \int s_A\big(U\big) s_B\big(U^{+}\big) [dU]_{\beta} }{ \int 1 \ [dU]_{\beta} } = \dfrac{ \int\limits_{-\pi}^{\pi} d\phi_1 \ldots \int\limits_{-\pi}^{\pi} d\phi_n \prod\limits_{a < b} \left| e^{i \phi_a} - e^{i \phi_b} \right|^{2\beta} \ s_A\big(e^{i \phi}\big) \ s_B\big(e^{- i \phi}\big)}{\int\limits_{-\pi}^{\pi} d\phi_1 \ldots \int\limits_{-\pi}^{\pi} d\phi_n \prod\limits_{a < b} \left| e^{i \phi_a} - e^{i \phi_b} \right|^{2\beta} }
\end{align}
\smallskip\\
which is the $\beta$-deformed unitary integral. The norm is

\begin{align}
\Big< J_A \Big| J_B \Big>_{n} = \int\limits_{n \times n} \dfrac{[dU]_{\beta}}{{\rm Vol}_{\beta}(n)} J_A\big[U\big] J_B\big[U^{+}\big] =
\delta_{AB} \ \dfrac{Q_A}{P_A} \ \dfrac{ [\beta n]_A }{[\beta n - \beta + 1]_A}
\end{align}
\smallskip\\
where ${\rm Vol}_{\beta}(n) = \int \ [dU]_{\beta}$, and $J_Y[U]$ denote the Jack polynomials of
power sums $s_k = \sum_a U_a^k$ of eigenvalues $U_a = e^{i \phi_a}$. In terms of the normalized
Jack polynomials, one has

\begin{align}
\int\limits_{n \times n} \dfrac{[dU]_{\beta}}{{\rm Vol}_{\beta}(n)} j_A\big[U\big] j_B\big[U^{+}\big] =
\delta_{AB} \ \dfrac{ D_A(n) }{ D_A(n - \delta) }
\end{align}
\smallskip\\
where $\delta = \dfrac{\beta-1}{\beta}$ is an often encountered shift.

\subsection*{The $\beta$-deformed Cauchy identity.}

A useful formula is the Cauchy-Stanley identity, sometimes also called the completeness condition:
\begin{align}
\exp\left( \sum\limits_{k} \dfrac{\beta}{k} s_k {\tilde s}_k \right) = \sum\limits_{R} \dfrac{P_R}{Q_R} J_R(s) J_R( {\tilde s} ) = \sum\limits_{R} j_R(s) j_R( {\tilde s} )
\end{align}
Just as many other formulas, this expansion looks simpler in terms of $j$-polynomials.

\subsection*{The $\beta$-deformed IZ integral.}

The character expansion of the $\beta$-deformed IZ integral has a form
\begin{align}
\int\limits_{n \times n} \dfrac{[dU]_{\beta}}{{\rm Vol}_{\beta}(n)} \int e^{\mbox{tr } (X U Y U^{+})} = \sum\limits_{R} \dfrac{P_R}{Q_R} \dfrac{J_R[X]J_R[Y]}{[\beta n]_A} = \sum\limits_{R} \dfrac{d_R}{D_R(n)} j_R[X] j_R[Y]
\end{align}
The only origin of $n$-dependence lies in the explicit factors $D_R(n)$ (equivalently, $[\beta n]_R$).

\subsection*{The Carlsson-Okounkov shift identity.}

The Carlsson-Okounkov identity has a form
\begin{align*}
\left< J_A\big( s_k - (-1)^k \dfrac{\beta - m - 1}{\beta} \big) \Big| J_B\big( s_k - (-1)^k \dfrac{m}{\beta} \big) \right> = \emph{}
\end{align*}
\begin{align}
\emph{} = \prod\limits_{(i,j) \in A} \Big( m + \beta ( A^{\prime}_j - i) + (B_i - j) + 1 \Big) \prod\limits_{(i,j) \in B} \Big( - m + \beta ( B^{\prime}_j - i) + (A_i - j) + \beta \Big)
\end{align}
The r.h.s. of this identity has a typical structure of denominators of Nekrasov functions.

\subsection*{5d deformations.}
All formulas in this Appendix possess direct generalization to the
McDonald polynomials, which depend on two parameters.
Generalizations of the IZ and BGW integrals can also be defined in
this way. The McDonald case complements $\beta$ by the "quantum-group"
$q$-deformation and, therefore, is relevant for the AGT relation between
$5d$ SYM theories and the conformal blocks of the "quantum" Virasoro
algebra. For some recent work in this direction, see \cite{5dJ}. In
principle, even further deformations and generalizations may exist,
for example, to the Askey-Wilson polynomials \cite{AW}. The role of
Askey-Wilson polynomials in the context of AGT relation remains to
be understood.

\section*{Appendix 2. Derivation of Ward identities for the Selberg model}

The Ward identities can be most simply derived by requiring the vanishing of the
integral of full derivative:
\begin{align}\label{76}
\int\limits_{0}^{L} dz_1 \ldots dz_N \sum\limits_{a = 1}^{N} \dfrac{\partial}{\partial z_a} z_a^{k_m+1} \left( \prod\limits_{i < j} (z_i - z_j)^{2\beta} \prod\limits_{i = 1}^{N} z_i^u (z_i - L)^v \ s_{k_1} \ldots s_{k_{m-1}} \right) = 0, \ \ s_k = \sum_i z_i^k, \ \ k_m > 0
\end{align}
where $L$ is an auxiliary parameter introduced for the purpose of derivation of the
Ward identities (we put $L=1$ in the end). Whenever possible, we substitute
the zero indices of the correlators with appropriate powers of $N$, in order not
to work with the $t_0$ variable. Differentiating the expression in the brackets, and
using for $k_m > 0$ the relations
\begin{align}
\sum\limits_{a = 1}^{N} z_a^{{k_m}+1} \dfrac{\partial}{\partial z_a} \prod\limits_{i < j} (z_i - z_j)^{2\beta} = \beta \left( - ({k_m}+1) s_{k_m} + \sum\limits_{p = 0}^{{k_m}} s_p s_{{k_m}-p} \right) \prod\limits_{i < j} (z_i - z_j)^{2\beta}
\end{align}
\begin{align}
\sum\limits_{a = 1}^{N} z_a^{{k_m}+1} \dfrac{\partial}{\partial z_a} \prod\limits_{i} z_i^u = u s_{k_m} \prod\limits_{i} z_i^u
\end{align}
\begin{align}
\sum\limits_{a = 1}^{N} z_a^{{k_m}+1} \dfrac{\partial}{\partial z_a} \prod\limits_{i} (z_i - L)^v = v \sum\limits_{a = 1}^{N} \dfrac{z_a^{{k_m}+1}}{z_a - L} \prod\limits_{i} (z_i - L)^v = \left( v \sum\limits_{h = 0}^{{k_m}} L^{{k_m} - h} s_h - L^{{k_m}+1} \partial_L \right) \prod\limits_{i} (z_i - L)^v
\end{align}
\begin{align}
\sum\limits_{a = 1}^{N} z_a^{{k_m}+1} \dfrac{\partial}{\partial z_a} s_{l} = l s_{{k_m} + l}
\end{align}
one finds
\begin{align*}
\Big( u + (k_m + 1)(1 - \beta) \Big) {\tilde C}_{k_1 \ldots k_m} +
\sum\limits_{i = 1}^{m-1} k_i {\tilde C}_{k_1 \ldots k_i + k_m \ldots k_{m-1}} +
\beta \sum\limits_{p = 0}^{k_m} {\tilde C}_{k_1 \ldots k_{m-1}, k_{m} - p, p} +\emph{}
\end{align*}
\begin{align}
\emph{} + v \sum\limits_{h = 0}^{k_m} L^{k_m - h} {\tilde C}_{k_1 \ldots, k_{m-1} h} -
L^{k_m} \big( L \partial_L \big) {\tilde C}_{k_1\ldots k_{m-1}} = 0
\end{align}
where
\begin{align}
{\tilde C}_{k_1 \ldots k_m}(N) = \int\limits_{0}^{L} dz_1 \ldots dz_N \prod\limits_{i < j} (z_i - z_j)^{2\beta} \prod\limits_{i = 1}^{N} z_i^u (z_i - L)^v \ s_{k_1} \ldots s_{k_m}
\end{align}
are the Selberg integrals with an additional parameter $L$. Actually, from now on this
parameter is not needed: it was only useful in derivation of the Ward identities.
Using the obvious homogeneity property
\begin{align}
L \partial_L {\tilde C}_{k_1 \ldots k_{m}} = \left( N + \sum\limits_{i = 1}^{m-1} k_i +
u N + v N + N(N-1)\beta \right) \ {\tilde C}_{k_1 \ldots k_{m}}
\end{align}
and putting $L = 1$, one finds that the original integrals $C_{k_1 \ldots k_m}(N) =
{\tilde C}_{k_1 \ldots k_m}(N)\Big|_{L=1}$ satisfy the Ward identities
\begin{align*}
\Big( u + v + 2 \beta N + (k_m + 1)(1 - \beta) \Big) C_{k_1 \ldots k_m} +
\sum\limits_{i = 1}^{m-1} k_i C_{k_1 \ldots k_i + k_m \ldots k_{m-1}} +
\beta \sum\limits_{p = 1}^{k_m-1} C_{k_1 \ldots k_{m-1}, k_{m} - p, p} + \emph{}
\end{align*}
\begin{align}
\emph{} + v \sum\limits_{h = 1}^{k_m-1} C_{k_1 \ldots k_{m-1}, h} - \left( N +
\sum\limits_{i = 1}^{m-1} k_i + u N + N(N-1)\beta \right) C_{k_1 \ldots k_{m-1}} = 0
\label{WardFinal}
\end{align}
Note that we explicitly moved the two contributions $\beta N C_{k_1 \ldots k_m}$
(which arise at particular values
of $p = 0$ and $p = k_m$ in the third term) into the first term. Similarly, we
explicitly moved the contributions
$v C_{k_1 \ldots k_m}$ and $v N C_{k_1 \ldots k_{m-1}}$ into the first and the last
terms, respectively. All these
trivial transformations are necessary to get rid of the zero indices in the correlators
and, hence, of presence of the $t_0$ variable in the partition function.

Because of the obvious formula
\be
\dfrac{1}{S} C_{k_1\ldots k_m} = \left. \left( \dfrac{1}{\beta} \dfrac{\partial}{\partial t_{k_1}} \right) \ldots \left( \dfrac{1}{\beta} \dfrac{\partial}{\partial t_{k_m}} \right) Z_S(t) \right|_{t = 0}
\ee
the same relations can be rewritten as differential equations known as generalized
Virasoro constraints:
\begin{align*}
\left[ \Big( u + v + 2 \beta N + (k + 1)(1 - \beta) \Big) \dfrac{\partial}{\partial t_k} + \beta \sum\limits_{m} m t_{m} \dfrac{\partial}{\partial t_{k + m}} + \sum\limits_{a + b = k} \dfrac{\partial^2}{\partial t_a \partial t_b} + v \sum\limits_{h = 1}^{k-1} \dfrac{\partial}{\partial t_h} \right] Z_S(t_1, t_2, \ldots ) =
\end{align*}
\begin{align}
= \beta \left( N + \sum\limits_{i = 1}^{m-1} k_i + u N + N(N-1)\beta \right) Z_S(t_1, t_2, \ldots ), \ \ \ k > 0
\end{align}
This completes the derivation of the Virasoro constraints for the Selberg model.

The trick (\ref{76}) with insertion
of a new dimensional parameter $L$ does not
work for the elliptic Selberg integral, as it
does not work in the original Dotsenko-Fateev
integrals: dimensionless parameters are present
in the both cases.
Still analogues of the Virasoro
constraints in these both cases exist,
they will be considered and analyzed elsewhere.

\end{document}